\documentclass[3p,times,procedia]{elsarticle}
\usepackage{nupha_ecrc}
\usepackage[utf8x]{inputenc}

\volume{00}

\firstpage{1}

\journalname{Nuclear Physics A}

\runauth{}

\jid{nupha}

\jnltitlelogo{Nuclear Physics A}

\usepackage{amssymb}
\usepackage[figuresright]{rotating}

\begin{document}

\begin{frontmatter}



%% Instructions from Editor: Please use the following \dochead only in the preprint version (e-print arXiv etc.); 
%% use empty \dochead{} when submitting to Nuclear Physics A!
\dochead{XXVIIth International Conference on Ultrarelativistic Nucleus-Nucleus Collisions\\ (Quark Matter 2018)}
%% e.g. \dochead{17th International Conference on Dynamical Processes in Excited States of Solids}

\title{Features of the IP-Glasma}


\author{Bj\"orn Schenke}
\author{Chun Shen}
\author{Prithwish Tribedy}
\address{Physics Department, Brookhaven National Laboratory, Upton, NY 11973, USA}

\begin{abstract}
We discuss differences between the IP-Glasma model and typical wounded-nucleon model like initial conditions. We point out that the IP-Glasma initial state is more compact in the transverse plane and produces a significant initial flow, both of which contribute to an increased radial flow in the subsequent hydrodynamic evolution. A larger bulk viscosity, compared to calculations that use other initial state models, is required to compensate for these effects and find agreement with experimental data. We further demonstrate the importance of the initial momentum anisotropy of the glasma for anisotropy measures in small collision systems such as p+Pb.
\end{abstract}


\end{frontmatter}

% \linenumbers

\section{Introduction}
Ultra-relativistic heavy ion collisions provide a unique opportunity to study many-body systems governed by quantum chromodynamics (QCD) in the laboratory.
One expects that a plasma of deconfined quarks and gluons is created in such collisions. Fluid dynamic simulations have been successfully used to describe a wide range of experimental observables in heavy ion collisions and have revealed their sensitivity to the equation of state, transport properties of the medium (such as shear and bulk viscosities), as well as early-time energy deposition and dynamics \cite{Heinz:2013th,Gale:2013da,deSouza:2015ena,Song:2017wtw}. 

The combination of a classical Yang-Mills initial state \cite{Schenke:2012wb,Schenke:2012hg}, viscous relativistic hydrodynamics \cite{Schenke:2010nt,Schenke:2010rr,Schenke:2011bn}, and a microscopic hadron cascade \cite{Bass:1998ca,Bleicher:1999xi} establishes a comprehensive framework for the dynamical description of ultra-relativistic heavy ion collisions, as long as boost-invariance is a good approximation.\footnote{The link between the Yang-Mills and hydrodynamic stages and potential isotropization in momentum space is the most uncertain aspect of the framework. Recent developments involving an intermediate effective kinetic theory stage are discussed in \cite{Kurkela:2018wud,Kurkela:2018vqr}.}
Previous event-by-event calculations within this setup have been very successful in reproducing (mid-rapidity) bulk observables and multi-particle correlation measurements in Pb+Pb collisions at the LHC and Au+Au collisions at RHIC \cite{Gale:2012rq,Schenke:2014zha}.
In particular, the IP-Glasma initial state has been superior to Monte-Carlo Glauber type models in that it was able to describe the $v_n$ distributions for all centralities \cite{Gale:2012rq,Schenke:2014zha}, the negative binomial fluctuations of the multiplicity \cite{Schenke:2012wb,Schenke:2012hg}, as well as details of the elliptic flow-multiplicity correlations in ultra-central events \cite{Schenke:2014tga}.

More recently, the same framework has been applied to smaller collision systems \cite{Dusling:2015gta}, such as p+p and p+A collisions \cite{Mantysaari:2017cni}. For very high multiplicity events, final state interactions are expected to play an important role, comparable to heavy ion collisions. However, initial state momentum correlations \cite{Schlichting:2016sqo,Mace:2018vwq,Mace:2018yvl} are also present and should grow in importance with decreasing multiplicity \cite{Greif:2017bnr}. Both contributions are included in the IP-Glasma + hydrodynamics calculation, with the initial anisotropy encoded in the (complete) energy momentum tensor used to initialize the hydrodynamic simulation. 

In this work we analyze two important features of the IP-Glasma initial condition, namely the larger radial flow it leads to compared to typical MC-Glauber type initial conditions (and consequently requires a larger bulk viscosity to produce agreement with experimental data), and the initial anisotropic flow, which becomes important in small collision systems.

\vspace{-0.2cm}

\section{Large radial flow}
One noticeable difference between hydrodynamic calculations using the IP-Glasma initial state \cite{Rose:2014fba,Ryu:2015vwa,Ryu:2017qzn} and other calculations (e.g. \cite{Bozek:2011wa,Weller:2017tsr,Bernhard:2018hnz}) is the need for a large bulk viscosity to entropy density ratio $\zeta/s$ to describe the experimentally observed mean transverse momentum $\langle p_T \rangle$. This is because of the larger radial flow generated when using the IP-Glasma initial state, which, in turn, is generated by a more compact initial geometry and the initial radial flow from the Yang-Mills stage of the evolution. The more concentrated hot spots of the IP-Glasma model do not play a significant role as they are washed out quickly in the viscous hydrodynamic evolution. This point has been quantified in \cite{Gardim:2017ruc}, where the effect of small scale structures, as present in the IP-Glasma initial state, was analyzed by smearing the initial energy density using cubic splines, which erase the small scale structure but keep the overall size of the system unchanged. 

In contrast, smearing with Gaussians also increases the system size, e.g.\,smearing with a Gaussian of width $1\,{\rm fm}$ increases the rms radius in 30-40\% central Pb+Pb collisions by approximately 15\% to a value comparable to that obtained with MC-Glauber initial conditions using an energy density deposition with Gaussians of width $\sigma=0.4\,{\rm fm}$.

The $\langle p_T\rangle$ is reduced by approximately 6.5\% when performing this smearing. Turning off the initial flow from the IP-Glasma, by setting $u^\mu$ to zero leads to another $\sim$10\% reduction of $\langle p_T\rangle$. In addition, differences in the functional form of the bulk-$\delta f$ correction to the distribution functions can play a role \cite{qm18poster}. 

We conclude that the compactness and initial flow in the IP-Glasma model play comparable roles in generating a larger radial flow than typical MC-Glauber like models, which explains the need for a larger bulk viscosity to entropy density ratio in IP-Glasma calculations. We also note that Bayesian analyses have constrained the initial geometry to be relatively compact as well. The geometric mean of the thickness functions of the two nuclei is consistently preferred by the data and generates more compact configurations than the usual wounded nucleon model (and similar eccentricities as the IP-Glasma) \cite{Bernhard:2016tnd,Bernhard:2018hnz}. 

\vspace{-0.2cm}

\section{Initial anisotropy}
As discussed above, the IP-Glasma model includes initial flow, which contains the typical momentum anisotropy of the color glass condensate framework \cite{Schlichting:2016sqo}, here computed in the dense-dense limit \cite{Schenke:2015aqa}.

It was analyzed in \cite{Greif:2017bnr} how the relative contribution from the initial anisotropy changes with multiplicity in a combined Yang-Mills + parton cascade framework. Here, we seek to perform a similar analysis in the combined Yang-Mills + viscous hydrodynamics framework. To do so, we evaluate the momentum space anisotropy, defined as $\epsilon_p = \sqrt{(\langle T^{xx}-T^{yy}\rangle^2+\langle 2 T^{xy}\rangle^2)/\langle T^{xx} +T^{yy}\rangle^2}$, 
where $T^{\mu\nu}$ are components of the energy momentum tensor in either the Yang-Mills or hydrodynamic calculation, and $\langle \cdot \rangle$ is the energy density weighted spatial average. 

\begin{figure*}[tb]
  \centering
  \begin{minipage}[t]{0.48\textwidth}
    \vspace{0pt}
    \includegraphics[width=\textwidth]{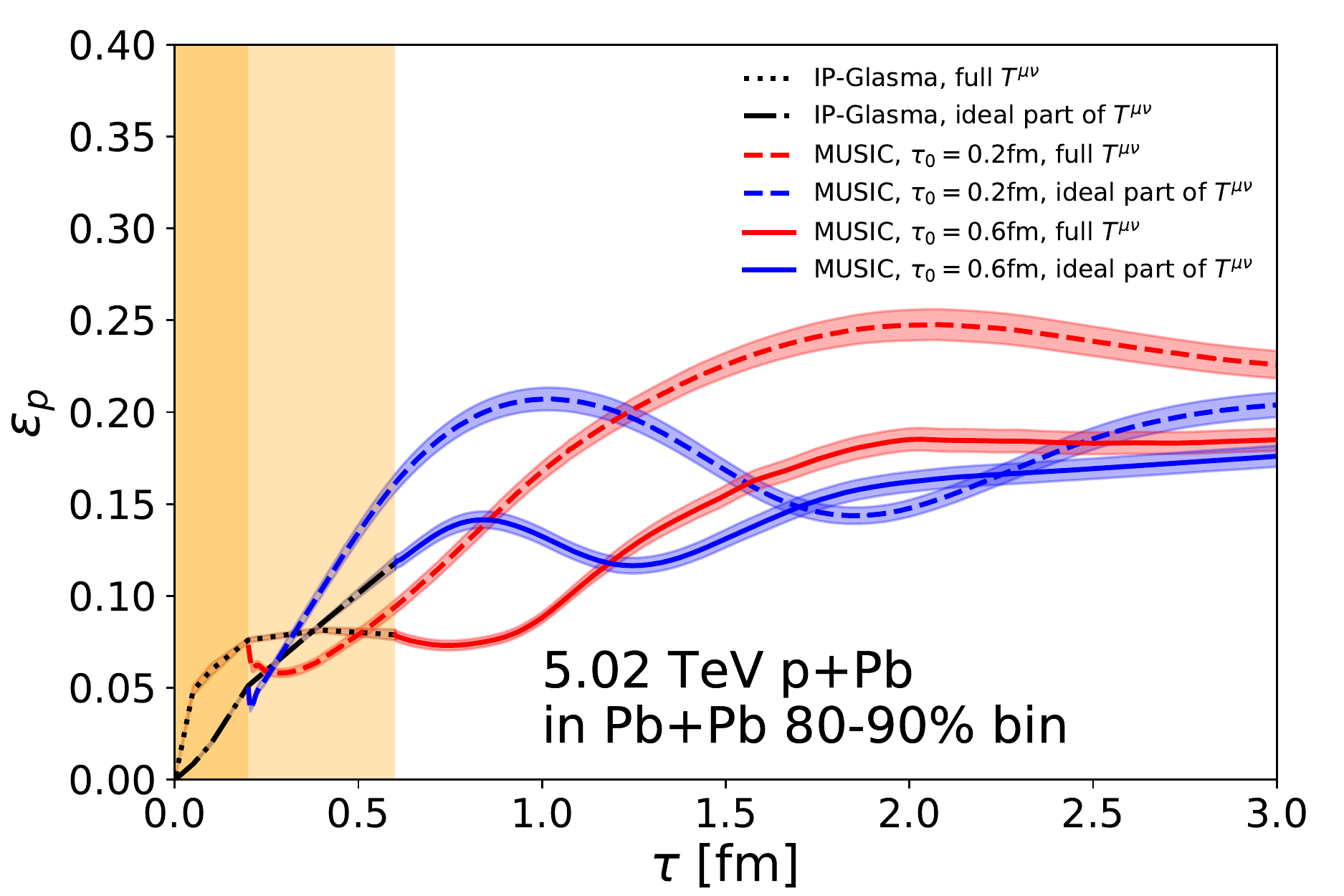} 
    \caption{Time evolution of the momentum anisotropy in the IP-Glasma and hydrodynamic stages of the evolution for two different switching times (0.2\,${\rm fm}$ (dashed) and 0.6\,${\rm fm}$ (solid)), using the full $T^{\mu\nu}$ (red) or its ideal part only (blue). \label{fig:aniso}}
  \end{minipage}
  \quad
  \begin{minipage}[t]{0.48\textwidth}
    \vspace{0pt}
    \includegraphics[width=\textwidth]{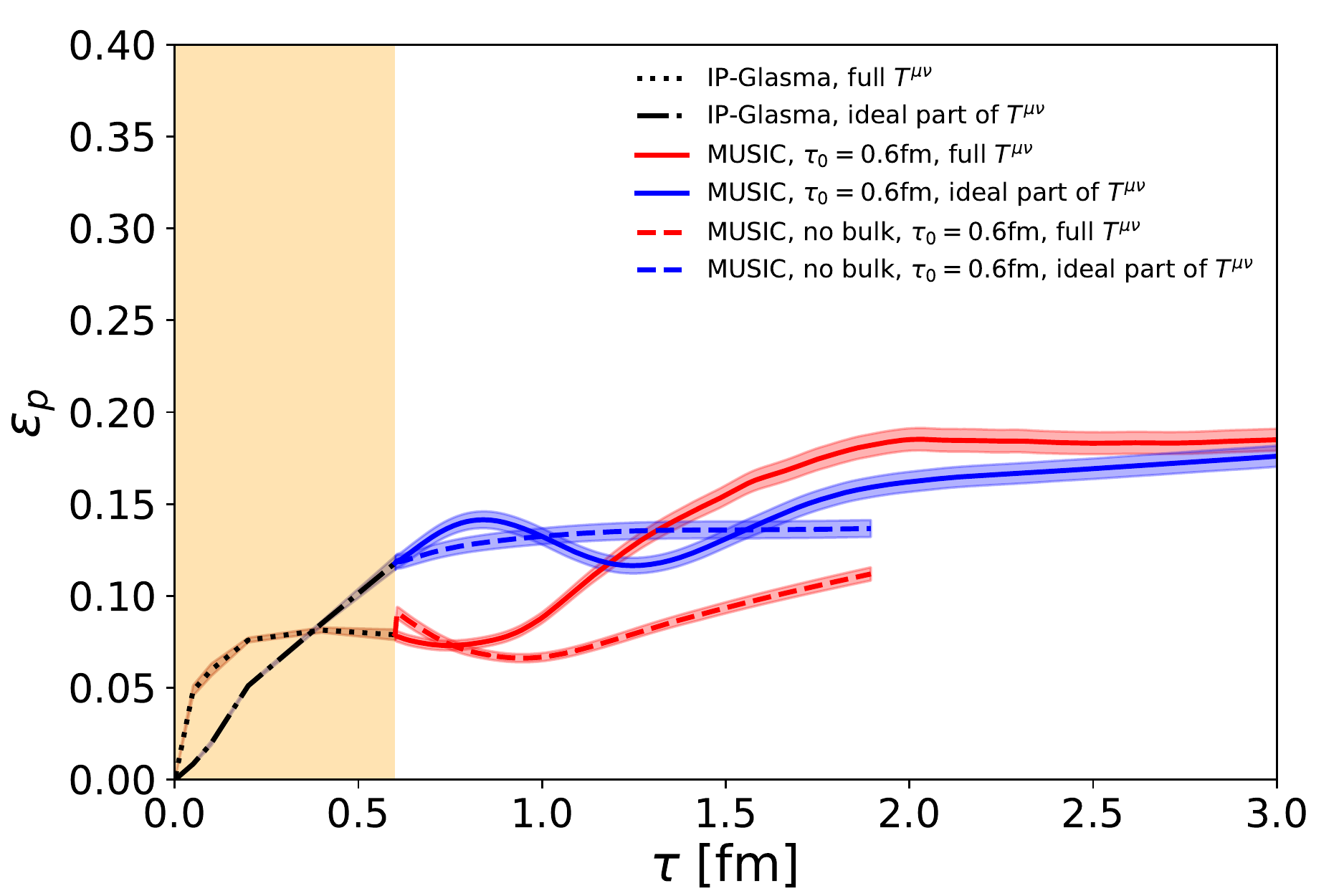} 
    \caption{Same as Fig.\,\ref{fig:aniso} for $\tau_{\rm switch}=0.6\,{\rm fm}$ showing results with (solid) and without (dashed) bulk viscosity. \label{fig:nobulk}}
  \end{minipage}
\vspace{-0.3cm}
\end{figure*}

For p+Pb collisions with multiplicities corresponding to those in 80-90\% central Pb+Pb collisions ($\langle dN_{\rm ch}/d\eta\rangle \approx 15$), Figure \ref{fig:aniso} shows the time evolution of the event averaged $\epsilon_p$, defined using either the full $T^{\mu\nu}$, or just its ideal part, which in the Yang-Mills sector is defined via the extracted energy density $\varepsilon$ and flow velocity as $T^{\mu\nu}_{\rm id} = (4/3)\varepsilon u^\mu u^\nu - (\varepsilon/3) g^{\mu\nu}$, using the ideal equation of state $\varepsilon = 3P$. Two choices for the switching time $\tau_{\rm switch}$ (0.2\,${\rm fm}$ and 0.6\,${\rm fm}$) are shown for comparison. In the very early times of the Yang-Mills evolution, the initial longitudinal fields decohere and the initial anisotropy becomes visible in $T^{\mu\nu}$. While $\epsilon_p$ comupted with the full $T^{\mu\nu}$ saturates to a constant value (which it retains due to the free streaming nature of the later stage Yang-Mills evolution), the ideal part keeps growing (as consequently does the viscous correction). 

After switching to hydrodynamics, we observe a brief initial decrease of the full $\epsilon_p$ (similar to what was seen in \cite{Greif:2017bnr}). Then hydrodynamics builds up more anisotropy (which is now correlated with the initial geometry). Comparing the relative values at the time of switching and after $3\,{\rm fm}$ evolution (when the values have approximately saturated), we find a 45\% initial state contribution when switching at 0.6\,${\rm fm}$ and 35\% when switching at 0.2\,${\rm fm}$. The sensitivity of the result to the precise matching time is a source of uncertainty that can hopefully be reduced by including a more sophisticated intermediate evolution stage (e.g. \cite{Kurkela:2018wud,Kurkela:2018vqr}).

Figure \ref{fig:nobulk} shows the effect of bulk viscosity. Results without bulk viscosity have a significantly shorter lifetime in the hydrodynamic phase, and therefore the contribution from initial flow increases to approximately 70\%. The initial jump in the result without bulk viscosity using the full $T^{\mu\nu}$ results from not matching the conformal equation of state (EoS) in the Yang-Mills phase to the lattice EoS in the hydrodynamic stage via an effective initial bulk contribution (which is done in all other scenarios).

\vspace{-0.2cm}

\section{Conclusions}
We have demonstrated that the presence of initial flow and the relatively compact initial geometry let hydrodynamical simulations initialized with the IP-Glasma framework produce more radial flow than typical MC-Glauber like models. This leads to the extraction of a larger bulk viscosity to entropy density ratio in the IP-Glasma framework. 

Furthermore, we have shown that the initial state anisotropy contained in the energy momentum tensor of the Glasma contributes significantly to the final state anisotropy in p+Pb events with multiplicities corresponding to 80-90\% central Pb+Pb events.

{\bf Acknowledgments} 
The authors are supported under DOE Contract No. DE-SC0012704. This research used resources of the National Energy Research Scientific Computing Center, which is supported by the Office of Science of the U.S. Department of Energy under Contract No. DE-AC02-05CH11231.



%% References

\vspace{-0.2cm}

\bibliographystyle{elsarticle-num}
\bibliography{spires.bib}






% \end{thebibliography}

\end{document}